\documentclass[a4paper,twosides]{article}
\usepackage{CJK,multicol,multirow,graphicx,fancyhdr,epstopdf}
\usepackage{amsmath,amsfonts,amssymb,bm,upgreek,mathrsfs,ccmap,mathcomp}
\usepackage[pagewise,switch,columnwise]{lineno}
\usepackage[compress,nospace]{cite}
\usepackage[dvipsnames]{xcolor}
\usepackage{CPL-2022}
\usepackage{threeparttable}

\newcommand{\cplyear}{2023} \newcommand{\cplvol}{39}
\newcommand{\cplno}{x} \newcommand{\cplpagenumber}{xxxxxx}
 \newcommand{\cplpage}{\cplpagenumber-\thepage}

\begin{document}

\begin{CJK}{GBK}{song}\vspace* {-4mm} \begin{center}
\large\bf{\boldmath{A search for radio pulsars in supernova remnants using FAST 
with one pulsar discovered}}
\footnotetext{\hspace*{-5.4mm}$^{*}$Corresponding author. Email: yanwm@xao.ac.cn

\noindent\copyright\,{\cplyear}
\href{http://www.cps-net.org.cn}{Chinese Physical Society} and
\href{http://www.iop.org}{IOP Publishing Ltd}}
\\[5mm]
\normalsize \rm{}Zhen Zhang$^{1,2}$, Wenming Yan$^{1,3,4*}$, Jianping Yuan$^{1,3,4}$, 
Na Wang$^{1,3,4}$, Juntao Bai$^{1,2}$, Zhigang Wen$^{1,3,4}$, 
Baoda Li$^{5}$, 
Jintao Xie$^{6}$, De Zhao$^{1,7}$, Yubin Wang$^{8}$ and Nannan Zhai$^{1,2}$
\\[3mm]\small\sl $^{1}$Xinjiang Astronomical Observatory, CAS, 150 Science 1-Street, 
Urumqi, Xinjiang 830011, China

$^{2}$University of Chinese Academy of Sciences, Beijing 100049, China

$^{3}$Key Laboratory of Radio Astronomy and Technology (Chinese Academy of Sciences), 
\\A20 Datun Road, Chaoyang District, Beijing, 100101, P. R. China

$^{4}$Xinjiang Key Laboratory of Radio Astrophysics, 150 Science 1-Street, Urumqi, 
Xinjiang 830011, China

$^{5}$GuiZhou University, Guizhou 550025, People's Republic of China

$^{6}$Research Center for Intelligent Computing Platforms, Zhejiang Laboratory, Hangzhou 311100, China

$^{7}$School of Physical Science and Technology, Xinjiang University, 
Urumqi, Xinjiang, 830046, China

$^{8}$School of Physics and Electronic Engineering, Sichuan University of Science \& Engineering, Zigong 643000, China.
\\[4mm]\normalsize\rm{}(Received xxx; accepted manuscript online xxx)
\end{center}
\end{CJK}
\vskip 1.5mm

\small{\narrower We report on the results of a search for radio pulsars in five 
supernova remnants (SNRs) with FAST. The observations were made using the 19-beam 
receiver in the Snapshot mode. The integration time for each pointing is 10 min. 
We discovered a new pulsar PSR J1845$-$0306 which has a spin period of 983.6 
ms and a dispersion measure of 444.6$\pm$2.0 cm$^{-3}$ pc in observations of SNR 
G29.6+0.1. To judge the association between the pulsar and the SNR, further 
verification is needed. We also re-detected some known pulsars in the data 
from SNRs G29.6+0.1 and G29.7$-$0.3. No pulsars were detected in observations 
of other three SNRs.

\par}\vskip 3mm
\normalsize\noindent{\narrower{PACS: 97.60.Gb, 97.60.Jd, 97.60.Bw}}\\
\noindent{\narrower{DOI: \href{http://dx.doi.org/10.1088/0256-307X/\cplvol/\cplno/\cplpagenumber}{10.1088/0256-307X/\cplvol/\cplno/\cplpagenumber}}

\par}\vskip 5mm
Before the discovery of pulsars, Baade \& Zwicky proposed that supernova 
explosions could produce neutron stars\ucite{bz34a}. Currently, it is generally 
believed that pulsars are fast-spinning, highly magnetized neutron stars. 
According to the standard model of pulsar evolution\ucite{go70}, pulsars are born in  
supernova explosions, so young pulsars with fast rotation are highly 
likely to be surrounded by SNR material. The two best examples 
supporting this model are the famous Vela pulsar and Crab pulsar. 
The Vela pulsar was discovered in 1968 at the edge of the Vela SNR\ucite{lvm68}. 
It is at the same distance and age as the SNR.
Also in 1968, the Crab pulsar was discovered at the center of the Crab 
Nebula\ucite{sr68}, 
which was subsequently shown to be associated with the Crab Nebula. 
The confirmation of the association of the Vela and Crab pulsars with the 
SNRs provides important observational evidence for the model 
of neutron stars produced by supernova explosions.

Finding pulsars in their parent SNRs is vital to study 
pulsar formation and the supernova explosion mechanism.
Attempts to search for pulsars in SNRs with large radio 
telescopes have been ongoing since the 1990s. Gorham et al.\ucite{gra+96} 
searched for pulsars in 18 SNRs using the Arecibo 305-m 
telescope at 430 and 1420 MHz, but did not find any new pulsars.
Kaspi et al.\ucite{kmj+96} conducted a pulsar search in 40 SNRs 
using the Parkes 64-m radio telescope at three frequencies of 436, 660 
and 1520 MHz, and discovered two new pulsars. 
Biggs et al.\ucite{bl96} used the Lovell 
76-m radio telescope to search for pulsars in 29 SNRs 
at 610 MHz, and found no new pulsars.
Lorimer et al.\ucite{llc98} searched for pulsars in 33 SNRs using 
the Lovell 76-m radio telescope at 606 MHz and discovered two new pulsars. 
Camilo et al.\ucite{clb+02} detected a 136-ms pulsar in SNR 
G54.1+0.3 using the Arecibo radio telescope. 
Gupta et al.\ucite{gmga05} used the GMRT radio telescope to discover a 
61.86-ms pulsar in SNR G21.5-0.9.
Zhang et al. (2018) employed the Parkes 64-m radio telescope to search for 
pulsars in SNR 1987A, but no new pulsars were detected. 
Straal and van Leeuwen\ucite{sv19} 
conducted a search for steep-spectrum pulsars in eight SNRs and 
pulsar wind nebulae using the LOFAR telescope, and discovered a pulsar 
candidate. Sett et al.\ucite{sbc+21} searched five SNRs for 
pulsars using the Green Bank radio telescope at 820 MHz, but no new 
pulsars were found. In the Pulsar Galactic Surface Survey (GPPS) 
project, Han et al.\ucite{hww+21} 
discovered several pulsars in three known SNRs using the FAST 
telescope. Further verification is needed to determine whether 
these pulsars are associated with SNRs. 

FAST has the unprecedented search sensitivity, it is expected to make 
breakthroughs in searching for pulsars in SNRs. In this Letter, 
we report on the search for radio pulsars in a sample of five SNRs 
(see Table 1 for details) with FAST. We select sources 
that have no associated pulsar or black hole and are located outside the 
sky coverage of two high priority pulsar searching projects CRAFTS\ucite{lwq+18} 
and GPPS. We also require that the angular size of the source is smaller 
than the field of view (FoV) of the Snapshot mode of FAST so that one 
Snapshot observation can completely cover the source. Based on these 
criteria, five SNRs in Table 1 are selected.

\begin{table*}
	\begin{center}
		\centering
		\begin{threeparttable}
			\caption{Properties of five SNRs\tnote{*}.}
		\begin{tabular}{lccccccccc}
			\hline
			SNR &RA  &DEC  &Size  &Distance &DM\tnote{**} & $T_{\rm sky}$ 
			&$T_{\rm rem}$ &Sensitivity
			\\
			&(hh:mm:ss) &(dd:mm:ss)  &(arcmin) &(kpc) &(cm$^{-3}$ pc) &K 
			&K & (mJy)\\
			\hline
			G29.6+0.1   &18:44:52 &$-$02:57:00 &5  &4.4$\sim$5   &315$\sim$475 &23 &12  & 0.02\\
			G29.7$-$0.3 &18:46:25 &$-$02:59:00 &3  &5.3$\sim$5.9 &559$\sim$759 &23 &210 &0.09 \\
			G30.7$-$2.0 &18:54:25 &$-$02:54:00 &16 &4.7          &210          &14 &1   &0.01\\
            G31.5$-$0.6 &18:51:10 &$-$01:31:00 &18 &6.3          &643          &19 &1   &0.02 \\       
            G96.0+2.0   &21:30:30 &+53:59:00   &26 &4$\sim$4.9   &180$\sim$226 &6  &1   &0.01\\

			\hline
		\end{tabular}
		\begin{tablenotes}
			\item[*] RA, DEC, Size and Distance are taken from the SNRcat\ucite{fs12} 
			and references therein.
			\item[**] DM is calculated using the YMW16 electron-density model\ucite{ymw17}.
		\end{tablenotes}
	    \end{threeparttable}
	\end{center}
\end{table*}  

{\it Observations and analysis.} 
Our data used in this work were obtained 
from the FAST 
observation project PT2020\_0050. The observations were 
made using the 19-beam L-band receiver of FAST in the Snapshot 
mode\ucite{hww+21} between 2021 February and May, which covers the 
1.0--1.5 GHz frequency band\ucite{jth+20}. All data from the 19 beams 
were recorded using the search-mode PSRFITS data format\ucite{hvm04} 
with a sampling time of 49.152 $\mu$s. The Snapshot observations last 
for 10 min for each pointing. Following Pan et al.\ucite{pqm+21}, 
we calculated 
the search sensitivities in Table 1 using the radiometer equation:
\begin{equation}
	S_{\rm min} = \frac{\beta (S/N_{\rm min}) T_{\rm sys}}
	{G \sqrt{n_{\rm p} t_{\rm int} \Delta f}}\sqrt{\frac{W}{P-W}}.
		\label{eq_sensitivity}
	  \end{equation}
In this expression, with our observing setup, the sampling efficiency 
$\beta$ is equal to 1, the width of the pulsar profile is $W$ = 10\% , 
the minimum signal-to-noise ratio is $S/N_{\rm min}$ = 10, 
the antenna gain $G$ = 16 K Jy$^{-1}$, 
the number of polarizations $n_p$ = 2, the integration time $t_{\rm int}$ 
is given in unit of seconds and the bandwidth $\Delta f$ = 
300 MHz in consideration of the radio-frequency interference (RFI), 
the system temperature $T_{\rm sys}$ consists of 
three components: the receiver noise $T_{\rm rcvr}$ = 24 K, the sky 
background noise $T_{\rm sky}$ which was calculated at 1250 MHz 
using the 408-MHz all-sky survey and a spectral index of $-$2.7, 
 and the contribution from the SNR radio radiation $T_{\rm rem}$. 
 Following the steps of Lorimer et al. (1998)\ucite{llc98}, we 
 estimated $T_{\rm rem}$ for each SNR and the results are given 
 in Table 1.

We performed periodicity and single pulse searches using the pulsar 
searching software package {\tt PRESTO}\ucite{ran01,rem02,rce03}. 
RFI was checked and zapped using the routine {\tt rfifind}. 
The range of trial dispersion measure (DM) was 0 -- 2000 
cm$^{-3}$ pc with a step of dmstep = 0.1. 
The routine {\tt accelsearch} was used to search periodic signals 
in the Fourier domain. We set $zmax$ = 200 to search for signals 
drifting up to 200 Fourier bins. The routine 
{\tt single\_pulse\_search.py} was used to search for single 
pulses with a signal-to-noise ratio larger than five. 

{\it G30.7$-$2.0, G31.5$-$0.6 and G96.0+2.0.} 
No pulsars were detected in these observations. However, we can obtain 
an upper limit (see Sensitivity in Table 1) on the pulsed 
radio emission from pulsars 
in these SNRs if there exist any pulsars in these regions.

{\it G29.7$-$0.3.} 
In observations of this SNR, we re-detected a known pulsar PSR 
J1845$-$0316 and a known Rotating Radio Transient (RRAT) 
PSR J1846$-$0257. The DM of PSR J1846$-$0257 is much smaller than that 
of G29.7$-$0.3, which implies that they are unlikely to be associated 
with each other. 
Although G29.7$-$0.3 and PSR J1845$-$0316 have similar DMs, 
it is difficult to unambiguously identify the association between them 
without accurate measurements for the distance to the pulsar and the 
remnant\ucite{lkm+00}.

{\it G29.6+0.1.} In observations of this SNR, we re-detected four known pulsars 
PSRs J1844$-$0244, J1844$-$0256, J1844$-$0302 and J1844$-$0310, and discovered a new 
pulsar J1845$-$0306. We checked with CRAFTS and 
GPPS people, but neither project has detected PSR J1845$-$0306. We also searched 
for this pulsar in the online Pulsar Survey 
Scraper\footnote[4]{https://pulsar.cgca-hub.org/search} which contains results 
of plenty of pulsar survey projects, but we did not find any pulsar with similar 
period and DM at the location of PSR J1845$-$0306. So we claim that PSR J1845$-$0306 
is a newly discovered pulsar in this work. 
PSR J1845$-$0306 was 
detected in the M11 beam in the third 
pointing and the M12 beam in the first pointing. 
The {\tt PRESTO} discovery plot is shown in Figure 1. 
The RA and DEC were estimated 
according to the centers of two beams in which the pulsar signal 
was detected. Then we fixed the RA and DEC, and fit for the period 
and DM only with the observational data using TEMPO2. The derived 
values are given in Table 2.
We also dedispersed and folded the data using these 
values and the {\tt DSPSR} package, 
and the result is presented in Figure 2. 
This pulsar has a rotation period of P = 983.6 ms and a dispersion 
measure of DM = 444.6$\pm$2.0 cm$^{-3}$ pc.  
Follow-up timing observations for this pulsar can determine the astrometric and spin 
parameters, such as position, spin period and DM more accurately. 
Gaensler et al.\ucite{ggv99} proposed that G29.6+0.1 is physically associated 
with the X-ray pulsar AX J1845$-$0258. However, the distance of AX J1845$-$0258 
is estimated to be $\sim$10 kpc which is significantly larger than that of 
G29.6+0.1. The DM of 
PSR J1845$-$0306 is in good agreement with that of G29.6+0.1.
Similar to G29.7$-$0.3 
and PSR J1845$-$0316, further investigations are needed to determine whether  
G29.6+0.1 is associated with PSR J1845$-$0306.

\begin{table}
	\begin{center}
		\centering
		\begin{threeparttable}
			\caption{The discovery parameters of the 983.6-ms pulsar.}
		\begin{tabular}{lc}
			\hline
			PSR  &J1845$-$0306 
			\\
			
			\hline
			RA (J2000) &18:45:25  \\
            Dec (J2000)   &$-$03:06:32 \\
            Spin period, P (ms)  &983.641 \\
            Timing epoch (MJD)   &59266  \\
			DM (cm$^{-3}$ pc)  &444.6$\pm$2.0 \\

			\hline
		\end{tabular}
	    \end{threeparttable}
	\end{center}
\end{table}  

{\it Discussion.} 
To date, nearly 4,000 
pulsars\footnote[2]{https://www.atnf.csiro.au/research/pulsar/psrcat/} 
and 383 
SNRs\footnote[3]{http://snrcat.physics.umanitoba.ca} 
in the 
Milky Way have been discovered. Among these, 110 pulsars have been 
identified to be associated with SNRs\ucite{czl+21}. That is, less 
than a third of known SNRs are observed to be in association with pulsars. 
The possible causes of the lack of detections of 
associated pulsars in SNRs are follows. Firstly, the radio emission beam of 
the pulsar may not sweep 
across the Earth, causing us to miss the pulsar. Secondly, the 
sensitivity limitation of the radio telescope may prevent us from 
detecting the pulsar\ucite{sv19}. Thirdly, the pulsar was born with 
a high kick velocity and the pulsar has moved out of the SNR. 
Fourthly, some models suggest that not all supernova explosions 
can produce neutron stars, or that the resulting neutron stars cannot 
be detected because they do not have the radio emission 
characteristics of radio pulsars\ucite{rs80,sbd84,man87,ns88}.

In this work, we did not detect any pulsar 
signals in observations of three SNRs G30.7$-$2.0, G31.5$-$0.6 
and G96.0+2.0. As the most sensitive single aperture radio telescope 
in the world, FAST can give the most stringent upper limit for the 
pulsed radio emission from the potential pulsars within these SNRs 
areas. 
We re-detected some known pulsars in observations of G29.7$-$0.3 
and G29.6+0.1, and discovered a new pulsar J1845$-$0306 
in observations of G29.6+0.1. The association between 
PSR J1845$-$0306 and SNR G29.6+0.1 is still unclear. On the one hand, PSR 
J1845$-$0306 is very close to the center of SNR G29.6+0.1, and they have 
consistent DMs, which support that they are associated, but on other hand, 
the spin period of PSR J1845$-$0306 is longer than that of 75 per cent pulsars 
associated with SNRs, which implies that PSR J1845$-$0306 is unlikely to be 
associated with a SNR. Judging associations between pulsars and SNRs depends on 
various criteria, such as consistency in distance and in age, 
positional coincidence, and evidence for interaction,
among which age and distance are the most fundamental. Follow-up 
timing observations for pulsars using FAST can help estimate their 
age and distance.

\begin{figure*}
	\includegraphics[width=\textwidth]{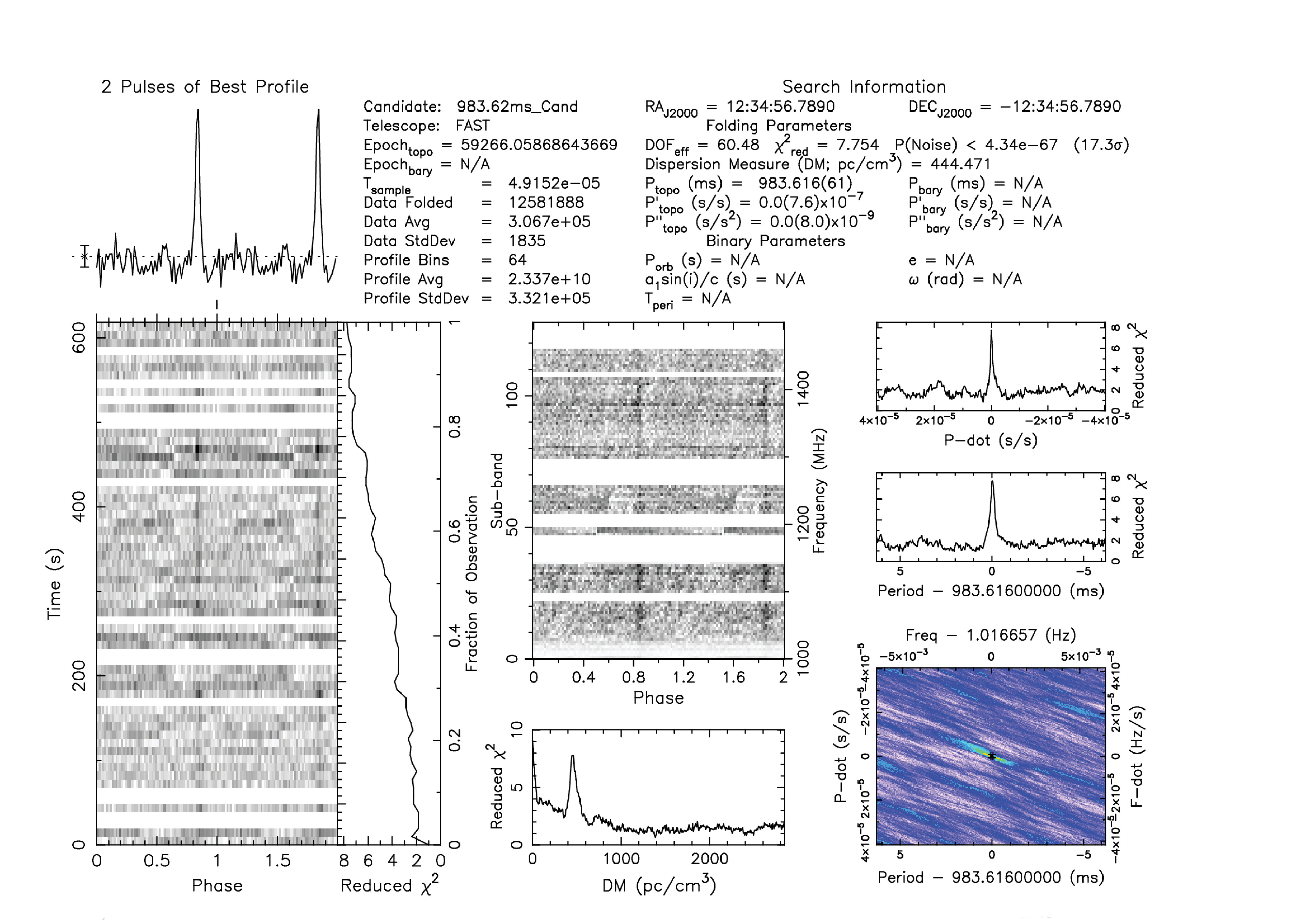}
	\caption{The {\tt PRESTO} discovery plot of the 983.6-ms pulsar.}
\end{figure*}

\begin{figure}
	\includegraphics[width=\columnwidth]{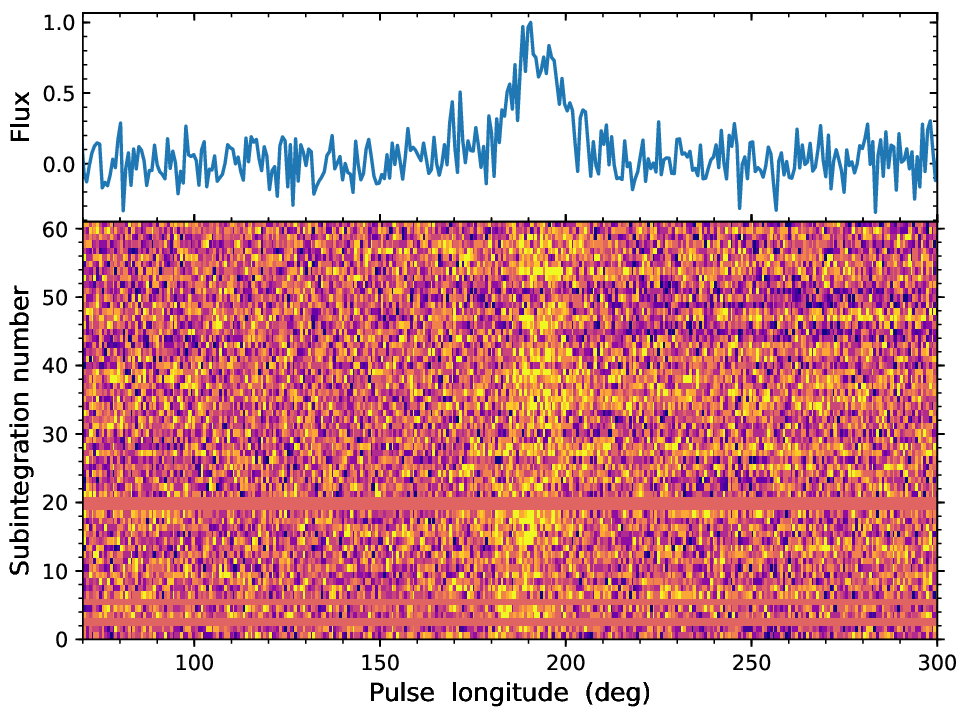}
	\caption{The pulse profile (upper) and single-pulse stack (lower) 
	folded using the {\tt DSPSR} package for the 983.6-ms pulsar.}
\end{figure}

\textit{Acknowledgements.} This work is supported by the National SKA 
Program of China (No. 2020SKA0120200), the National Natural Science 
Foundation of China (NSFC) 
project (No. 12041303, 12273100, 12041304, 12288102), 
the National Key R\&D Program of China (No. 2022YFC2205201), 
the West Light 
Foundation of Chinese Academy of Sciences (No. WLFC 2021-XBQNXZ-027), 
the Major Science and Technology Program of Xinjiang Uygur 
Autonomous Region (No. 2022A03013-4), the Natural Science Foundation 
of Xinjiiang Uygur Autonomous Region (No. 2022D01D85), and the open 
program of the Key 
Laboratory of Xinjiang Uygur Autonomous Region No. 2020D04049. 
This research is partly supported by
the Operation, Maintenance and Upgrading Fund for Astronomical 
Telescopes and Facility Instruments, budgeted from
the Ministry of Finance of China (MOF) and administrated
by the CAS. This work made use of the data from FAST 
(Five-hundred-meter Aperture Spherical radio Telescope). FAST is 
a Chinese national mega-science facility, operated by National 
Astronomical Observatories, Chinese Academy of Sciences.


\end{document}